\documentclass[iop]{emulateapj}

\usepackage{textcomp}
\usepackage{array,multirow}
\usepackage{natbib}
\usepackage{amsmath}
\usepackage{xcolor}
\defcitealias{KH09}{KH09}

\begin{document}

\title{Do You See What I See? Exploring the Consequences of Luminosity Limits in Black Hole-Galaxy Evolution Studies}







\author{Mackenzie L. Jones\altaffilmark{1}, Ryan C. Hickox\altaffilmark{1}, Simon J. Mutch\altaffilmark{2}, Darren J. Croton\altaffilmark{3}, Andrew F. Ptak\altaffilmark{4}, Michael A. DiPompeo\altaffilmark{1}}
\affil{$^{1}$Department of Physics and Astronomy, Dartmouth College, Hanover, NH 03755, USA}
\affil{$^{2}$School of Physics, University of Melbourne, Parkville, Victoria 3010, Australia}
\affil{$^{3}$Centre for Astrophysics \& Supercomputing, Swinburne University of Technology, PO Box 218, Hawthorn, VIC 3122, Australia}
\affil{$^{4}$NASA Goddard Space Flight Center, Code 662, Greenbelt, MD 20771, USA}

\begin{abstract}

In studies of the connection between active galactic nuclei (AGN) and their host galaxies there is widespread disagreement on some key aspects stemming largely from a lack of understanding of the nature of the full underlying AGN population. Recent attempts to probe this connection utilize both observations and simulations to correct for a missed population, but presently are limited by intrinsic biases and complicated models. We take a simple simulation for galaxy evolution and add a new prescription for AGN activity to connect galaxy growth to dark matter halo properties and AGN activity to star formation. We explicitly model selection effects to produce an ``observed'' AGN population for comparison with observations and empirically motivated models of the local universe. This allows us to bypass the difficulties inherent in many models which attempt to infer the AGN population by inverting selection effects. We investigate the impact of selecting AGN based on thresholds in luminosity or Eddington ratio on the ``observed'' AGN population. By limiting our model AGN sample in luminosity, we are able to recreate the observed local AGN luminosity function and specific star formation-stellar mass distribution, and show that using an Eddington ratio threshold introduces less bias into the sample by selecting the full range of growing black holes, despite the challenge of selecting low mass black holes. We find that selecting AGN using these various thresholds yield samples with different AGN host galaxy properties.

\end{abstract}


\section{Introduction}\label{sec:intro}

In the past decade there has been significant progress in building a generalized model of the formation and evolution of galaxies and their host halos across cosmic time (See \citealt{Sil12} for a review). These galaxy formation models have illustrated the impact of stellar and AGN feedback on galaxy growth (e.g., \citealt{Kau93,Spr05feed,Bow06,Cro06,Gen14,EAGLE,Vol15,Kha15,Fen16}). However, the connection between galaxy growth and black hole growth is not well understood. The volume-averaged galaxy-black hole growth rate is consistent with the black hole-spheroid mass relationship (e.g., \citealt{Hec04}) and the masses of supermassive black holes (SMBH) are found to be correlated with host stellar bulge properties \citep{Kor13}, however the physical processes connecting black holes to their galaxies are still uncertain \citep{Ale12}.

One major challenge lies in understanding selection effects (e.g., color selection, obscuration, flux limits, and luminosity limits) in observed AGN samples. Selection effects become apparent as different methods across wavelengths intended to probe the same physical phenomena yield populations with different AGN and galaxy properties. It is difficult to correct for selection effects as it is not immediately apparent which effects, or even how many, are influencing what is observed.
For example, the distribution of the Eddington ratio (the ratio of bolometric luminosity to the Eddington limit) in the X-rays is observed to be approximately power-law in shape and spans several orders of magnitude, while in the optical, the Eddington ratio distribution is observed to vary by galaxy age \citep{KH09}. This discrepancy between observed accretion rate distributions, where accretion rate refers to the specific accretion rate, i.e. Eddington ratio,  can be resolved when selection effects in spectroscopic classification are taken into account \citep{Jon16}. 
Additionally, AGN are preferentially found in the most massive, bulge dominated galaxies across all wavelengths. X-ray selected AGN, in particular, are observed to reside in more massive host dark matter halos compared to optical AGN (e.g., \citealt{Hic09,Kou13,Ric13}). Furthermore, the probability of finding an AGN is tied to the host galaxy star formation rate: the likelihood of finding an AGN increases for star forming galaxies compared to quiescent galaxies for a given accretion rate (e.g., \citealt{Aza15}). Yet the observation that AGN preferentially reside in massive galaxies is likely an effect due to AGN selection above a certain luminosity threshold. For a given luminosity threshold, it is possible to probe lower in Eddington ratio for high mass AGN compared to low mass AGN, for which only the highest accreting systems would be selected \citep{Air12}.
The strength of AGN clustering measurements are also shown to vary across wavelengths (e.g., \citealt{Hic09,Men16}). 

In recent years, there has been increased insight into the physical origin of the black hole accretion rate distribution. This distribution is expected to cover a wide dynamic range and has converged on a broad universal shape that may be dependent on star formation (e.g., \citealt{Air12,Bon12,Che13,Hic14,Aza15}). Further evidence for this broad distribution is presented by \citet{Vea14} in which a variety of accretion distributions are used to recreate the observed quasar luminosity function (QLF). An individual AGN may vary on short timescales within the dynamic range of this AGN accretion rate distribution (e.g., \citealt{Hic14,Sch15}), while galaxies vary on longer timescales, on order of $\gtrsim100$ Myr (e.g., \citealt{Ale12}). This variability may be due to accretion disk instabilities, or feedback on the accreting material (e.g., \citealt{Sie97,Hop05,Jan11,Nov11}), however, most observational evidence of this variability is indirect (e.g., \citealt{Sch10qso,Kee12,Sch13,Sar16}). Due to this variability (or ``flickering'') each observation is a snapshot of that source's current AGN accretion: an instantaneous value, rather than a description of the average black hole activity. 

Further attempts have been made to understand the interplay between AGN and their host galaxies through theoretical modeling. Hydrodynamical models, in particular, investigate AGN accretion and explicitly spatially resolve feedback on the interior halo of the host galaxy. Recent models, such as \textsc{eagle} \citep{EAGLE}, \textsc{illustris} (\citealt{Vog14,Gen14}), \textsc{massiveblack-II} \citep{Kha15}, and BlueTides \citep{Fen16} are built to simultaneously describe complex physical processes and dark matter halo growth providing spatial and phase information, although this usually involves sub-grid modeling and many free parameters. These hydrodynamical simulations have had success in modeling a variety of AGN properties (e.g., winds, gas properties, luminosity functions), but number statistics are sacrificed for better resolution due to the high computational cost. Alternatively, semi-analytic models separate baryonic physics from dark matter halo growth and sacrifice spatial information by averaging over the spatial scale which makes them less computationally expensive than hydrodynamic models. This also usually involves utilizing many free parameters that are often degenerate, such that model predictions are not unique and can be difficult to interpret (e.g., \citealt{Hen09,Lu11,Mut13sam}). 

In order to investigate the impact of selection effects on AGN observations, we have built a simple semi-numerical model for galaxy evolution and AGN accretion based on the observed connections between AGN and their host galaxies. In this work we take a ``forward'' modeling approach in which we simulate an intrinsic full galaxy population with a complete knowledge of its physical properties and apply known limits due to selection to compare to observations, rather than modeling unknown selection effects in observations. We combine the best characteristics of a semi-analytic simulation, a fast model with a low number of parameters, and the expository power of a straightforward, intuitive prescription for galaxy and black hole growth. 
We discuss our method for forward modeling the local AGN population in Section \ref{sec:method}. This is broken into two subsections that discuss the galaxy formation model (\ref{ssec:gal}) and AGN prescription (\ref{ssec:agn}). In this paper we focus on the most simple of selection effects, namely luminosity limits that are primarily driven by the sensitivity limits of X-ray surveys. We compare our simulated AGN population to observations by imposing limits on luminosity and Eddington ratio in Section \ref{sec:agndef}. A discussion of our results and summary are given in Section \ref{sec:sum}.

We utilize the same assumed cosmology as in \citet{Mut13}, in which a 1-year Wilkinson Microwave Anisotropy Probe (WMAP1; \citealt{Spe03}) cold dark matter (CDM) cosmology with $\Omega_\text{m}=0.25$, $\Omega_\Lambda=0.75$, and $\Omega_\text{b}=0.045$ is used. Likewise, all results are shown with a Hubble constant of $h=0.7$, where $h\equiv H_0/100\;\text{km s}^{-1}\;\text{Mpc}^{-1}$.

\section{Method of Simulation}\label{sec:method}

\subsection{The Semi-Numerical Galaxy Formation Model}\label{ssec:gal}

The host galaxies of our model AGN sample are built from a simulation of galaxies and dark matter halos using the formation history model of \citet{Mut13}. This model begins with a sample of dark matter halos from the N-body Dark Matter Millennium Simulation \citep{Spr05}, where the baryonic content of these halos is determined by the halo growth rate and the cosmological fraction of baryons compared to dark matter. The Millennium Simulation uses $N=10^{10}$ particles in a volume of $714\times714\times714$ Mpc and follows galactic evolution from $z=127$ to $z=0$ in 64 time steps of $\sim200-350$ Myr.

\citet{Mut13} combines two simple functions to connect galactic growth to the formation history of the Millennium Simulation dark matter halos; a baryonic growth function and a physics function. The baryonic growth function regulates the availability of baryonic material to be used by stars, mapping the dark matter growth history to the star formation rate (SFR). The physics equation represents the combined effects of internal and external physical processes (e.g. shock heating, supernova feedback, AGN feedback, galaxy mergers, tidal stripping, etc.) on the star formation rate, acting as an efficiency of baryonic matter consumption. 

In this model, the conversion efficiency of the baryons into stars is based on a simple function (Equation \ref{equ:simphy}) that represents baryon cooling, star formation, and both stellar and AGN feedback physics as one net component, rather than treating them individually which can be computationally costly or can introduce significant uncertainty due to a large number of physical parameters: 
\begin{equation}\label{equ:simphy}
M_{*}/M_{}=\mathcal{E}_{M_{vir}}\exp\left(-\left(\frac{\Delta M_{vir}}{\sigma_{M_{vir}}}\right)^2\right),
\end{equation}
where $\mathcal{E}_{M_{vir}}$ represents the baryonic matter conversion efficiency and $M_{vir}$ is the halo virial mass with standard deviation $\sigma_{M_{vir}}$. Treating the baryonic matter in this way allows for each galaxy's formation to follow the growth history of its individual halo, capturing the true diversity of galaxy formation histories, rather than adding an artificial scatter as is done with many other methods. This approach has proven to be accurate for $z<2$ in reproducing the derived star formation rates as a function of halo mass (\citealt{Con09,Bet12}). The simplicity is also important and makes the model general and flexible such that the functional form can be used for either the virial mass of the halo (M$_{vir}$) or the instantaneous maximum circular velocity (V$_{max}$), which is more directly tied to the gravitational potential. Both M$_{vir}$ and V$_{max}$ provide good fits and can be used interchangeably, although V$_{max}$ may be more tightly coupled with the stellar mass growth \citep{Red13}. 

The \citet{Mut13} model is able to accurately recreate the local stellar mass function and its evolution to $z\sim3$. It also has the capability to trace the stellar mass function as a function of stellar age, which allows for passive and star forming galaxies to be treated separately. For more information on this model, please refer to \citet{Mut13}.

In the \citet{Mut13} semi-numerical model, galaxies are assigned a star formation rate (SFR) following the growth of the dark matter halos and availability of the baryonic material. If the galaxy is not flagged as the central galaxy of a dark matter halo, it is assigned SFR$\;=0$. Since these leftover ``satellite'' galaxies are passive with low SFR, they may be treated with this simplification when investigating the local and global stellar mass function. However, in this work we are directly comparing to the observed SFR distribution and this simple assumption of assigning ``satellite'' galaxies a SFR$\;=0$ is no longer applicable. There are also a fraction $\sim59\%$ of central galaxies with SFR$\;=0$ for which we calculate a SFR.

We have devised a simple prescription for repopulating galaxies with SFR$\;=0$ based on the distribution of specific star formation rate (sSFR; SFR/M*) of our simulated passive central galaxies. For central galaxies with SFR$\;=0$, the SFR are smoothed over three redshift bins in order to limit the number of central galaxies with SFR$\;=0$. Any galaxy with SFR$\;=0$, including central galaxies with SFR$\;=0$ after the smooth, are assigned a sSFR representative of a passive galaxy. The distribution of sSFR for central galaxies is well fit by two gaussians, so we assign an sSFR consistent with the passive gaussian distribution. Using this simple method, we can repopulate the SFR for our ``satellite'' and central galaxies in order to more accurately compare to observations. The repopulated sSFR drawn from the passive central galaxy sSFR distribution are low enough that they do not greatly impact the average doubling time for our simulated sample (e.g. for a stellar mass of $\sim10^{10}\;M\odot$ the typical sSFR assigned is $10^{-11}\;\text{yr}^{-1}$, corresponding to a doubling time of $\sim100$ Gyr, and as such, the corresponding increase in the stellar mass at each time step would be negligible).

\subsection{A Simple Prescription for AGN Accretion}\label{ssec:agn}

Since the \citet{Mut13} galaxy evolution model synthesizes complicated physics down to a one-dimensional function, we are able to add complexity in the form of an AGN component and limit the number of free parameters, while remaining computationally tractable. The foundation of our simple prescription for AGN accretion is motivated by the work of \citet{Jon16}, in which it is assumed the instantaneous observed AGN luminosity is due to short-term variability (e.g., \citealt{Ale12,Hic14}) and follows an Eddington ratio ($\text{L}_{bol}/\text{L}_{Edd}$) distribution described by a Schechter function, a power law with an exponential cutoff near the Eddington Limit \citep{Hop09}:
\begin{equation}\label{equ:sch}
\frac{dt}{d\log \text{L}_{bol}}= \left(\frac{\text{L}_{bol}}{\text{L}_{Edd}}\right)^{-\alpha}\exp\left(-\text{L}_{bol}/\text{L}_{\rm{Edd}}\right).
\end{equation}
An Eddington ratio distribution that takes the functional form of a Schechter function consists of only two free parameters: the slope of the power law $\alpha$, and the lower cut off $\text{L}_{cut}$ which sets the amplitude of the function such that the integral of the curve is one.

Our black hole masses are derived from the black hole-bulge relationship of \citet{Har04}, with total galaxy stellar mass from the model galaxies used as a proxy for bulge mass \citep{Kor13}. We note that there is significant uncertainty and scatter in the relationship between black hole mass and total stellar mass, with marked differences observed for local ellipticals and AGN hosts (e.g., \citealt{Rei15}). We adopt the  \citet{Har04} relationship as it approximately bisects that found for ellipticals and AGN hosts, and broadly represents the correlation observed for the full galaxy population. 
Furthermore, we do not include a redshift evolution in our black hole mass calculations. There is widespread disagreement about whether a redshift evolution is present or if any observed evolution is caused by a sample selection bias (e.g. \citealt{Dec10,Mer10,Scz11,Ben11,Woo13,Deg15,Sha16}). We tested the impact of a moderate evolution \citep{Mer10} on our analysis and found that our simulated luminosity functions for both an evolving and non-evolving black hole-galaxy relationship were consistent within their random scatter for redshifts up to $z\sim3.5$.

For simplicity, we assume a constant black hole-galaxy relationship. We can thus calculate the Eddington Luminosity, $\text{L}_{Edd}=1.38\times10^{38}\;\text{M}_{BH}$. We then select an instantaneous AGN bolometric luminosity from the Eddington ratio distribution. Both the galaxy and AGN components are defined in terms of bolometric luminosities; using scaling relationships these can be converted into different broad-band luminosities. In this paper we focus on the hard ($2-10$ keV) X-rays, although we note that this approach could be equally valuable in the $>10\;\text{keV}$, soft X-rays, infrared, and/or optical regime.

Galactic X-ray emission is produced by a combination of high mass X-ray binaries (HMXB), low mass X-ray binaries (LMXB), and hot gas (e.g., \citealt{Hor05,Leh10,Fra13}). In this work we ignore hot gas since X-ray binaries dominate the X-ray emission above $1.5\;\text{keV}$ \citep{Leh16}. We use the \citet{Leh16} scaling relationships given by
\begin{align}\label{equ:xrb}
\text{L}_{x,LMXB}(z)&=\alpha_0(1+z)^\gamma M_* ,\nonumber \\
\text{L}_{x,HMXB}(z)&=\beta_0(1+z)^\delta SFR ,
\end{align}
where $\log\alpha_0=29.30\pm0.28$, $\gamma=2.19\pm0.99$, $\log\beta_0=39.40\pm0.08$, $\delta=1.02\pm0.22$, and the contribution from HMXB scales with star formation rate (SFR) while LMXB scales with stellar mass.

The AGN X-ray luminosity is derived from the bolometric AGN luminosity using the \citet{Lus12} observed relationship for Type 1 AGN,
\begin{equation}
\log \text{L}/\text{L}_{[2-10 keV]}=(m\pm dm)\text{L}_{Edd}+(q\pm dq),
\end{equation}
where $m\pm dm=0.752\pm0.035$ and $q\pm dq=2.134\pm0.039$.
We use the function for Type 1 AGN to model the implicit relation for our full sample. Rather than utilizing the \citet{Lus12} Type 2 AGN relationship for an ``obscured'' percentage of our full sample, we add in obscuration using a different method to lessen the influence of potential observational biases incurred in observing Type 2 AGN. 

We randomly assign some of our model AGN to be obscured based on the relationship between L$_X$ and the obscured fraction from \citealt{Mer14} (See their figure 9). For AGN that are ``unobscured'', we assign a column density of $\log N_{H}=20\;\text{ cm}^{-2}$, while those selected to be ``obscured'' have column densities drawn from the NuSTAR-informed $N_{H}$ distribution (\citealt{Lan15}; Figure 13b). Based on these column densities and X-ray spectral models (e.g., \citealt{Lan14}) we then add X-ray absorption to our obscured sample. We run our simulation both with and without this prescription for obscuration and find that adding obscuration does not significantly alter the results of our analysis, other than extending the lower cutoff of the AGN X-ray luminosities from $\text{L}_X=10^{37.5}\;\text{erg s}^{-1}$ to $\text{L}_X=10^{35.5}\;\text{erg s}^{-1}$ (with obscuration) and altering the amplitude of our AGN X-ray luminosity function (see Section \ref{ssec:luminosity}). There is no change for the highest luminosities ($\text{L}_X>10^{44}\;\text{erg s}^{-1}$). We thus have a simulated galaxy and AGN X-ray luminosity for every object in our sample and can investigate the properties of this X-ray emission to compare to observations.

\section{Comparison of the Simulated AGN Population with X-ray Observations}\label{sec:agndef}

In our simulation we built bolometric AGN and galaxy luminosities which are then scaled into X-ray luminosities. At this point our X-ray sample represents an intrinsic snapshot of the full AGN population. In order to compare our simulation to observations we must forward model the selection effects we would expect to see in these observed results. 

In the literature, there are a range of different parameters used to define an AGN. In the X-ray band, AGN are most often selected based on an observed luminosity threshold. This threshold is often set at the transition where AGN emission begins to dominate host galaxy emission ($\text{L}_X\sim10^{41.5}\;\text{erg s}^{-1}$). Despite the ease of this method for selecting the brightest AGN, by definition it does not select less luminous AGN, especially when these AGN have luminosities comparable to their host galaxy (e.g., highly star forming galaxies). 

Alternatively, an AGN may be defined by its accretion rate using colors, broad lines, or by the Eddington ratio. Typically, this threshold is set at an Eddington ratio of $\lambda\gtrsim0.01$ (e.g., \citealt{Hop09}). 
Since the Eddington ratio is closely linked to the black hole mass, selecting an AGN in this way often relies on accurate measurements of the black hole, which may be difficult to acquire for large samples. Additionally, it becomes challenging to separate AGN accretion from other accretion processes at low luminosities (e.g., \citealt{Leh16}). An Eddington ratio limit, however, is better for selecting low-luminosity AGN compared to a luminosity limited sample, as demonstrated by our following analysis. In this work we are able to fully explore this threshold since each of our simulated AGN has a known Eddington ratio drawn from our Schechter function distribution.

\subsection{The AGN Luminosity Function}\label{ssec:luminosity}

We first compare our model AGN population to the observed AGN X-ray luminosity function (XLF). This distribution is particularly useful for investigating AGN evolution but observations are uncertain at faint luminosities and high redshifts due to poor number statistics and contamination (e.g., \citealt{Nan05,Air10}). To mitigate these problems, luminosity limits are commonly placed on samples, thereby adding a known selection effect.

\subsubsection{The AGN luminosity function at $z\;=0$}

\begin{figure}[!t]
\resizebox{85mm}{!}{\includegraphics{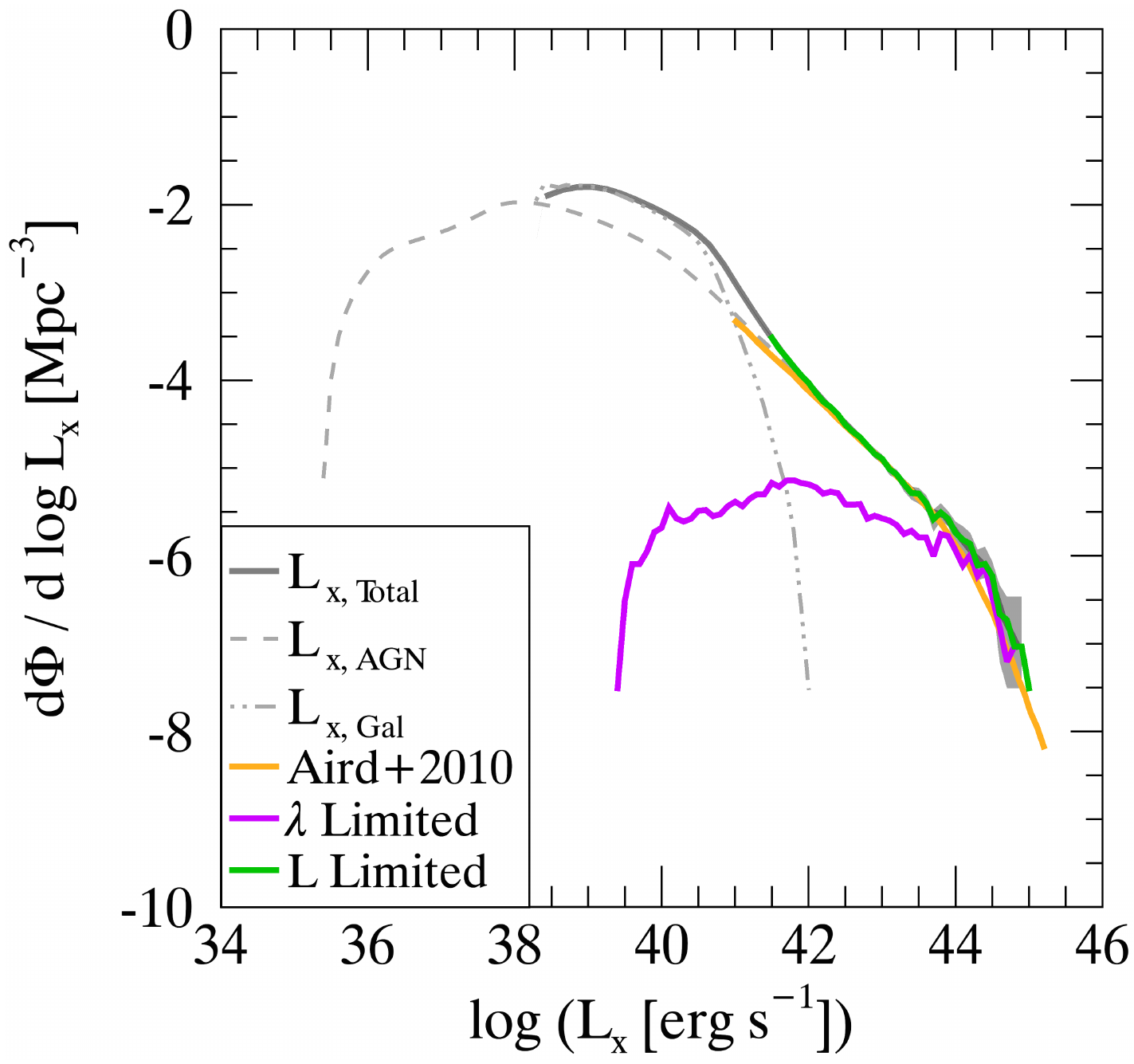}} \\
\caption{Comparison of the X-ray luminosity function (XLF) from our model at $z\;=0$ to results from X-ray observations. The XLF for the full sample of model galaxies is given as a solid grey line while the galaxy and AGN components that make up the total X-ray emission are shown as light grey triple-dot-dash and dash lines, respectively. The turnover of the total X-ray luminosity at $\text{L}_X\sim10^{38}\;\text{erg s}^{-1}$ is caused by the minimum contribution from star formation as defined by the mass limit of our model sample. The model XLF is then compared to the distributions of the luminosity and Eddington ratio limited model samples: luminosity limit of $\text{L}_X\sim10^{41.5}\;\text{erg s}^{-1}$ (green) and Eddington ratio limit of 0.01 (magenta). We further compare our limited model samples to the XLF of \citet{Air10} at $z=0$ (orange). We find that our luminosity limited sample most closely matches the full theoretical curve. While at high luminosities ($\text{L}_X\gtrsim10^{44}\;\text{erg s}^{-1}$) we are able to match the \citet{Air10} XLF with our full sample XLF and both of our limited sample XLF. \label{fig:lfn}}
\end{figure}

\begin{figure*}[!]
\begin{center}
\resizebox{150mm}{!}{\includegraphics{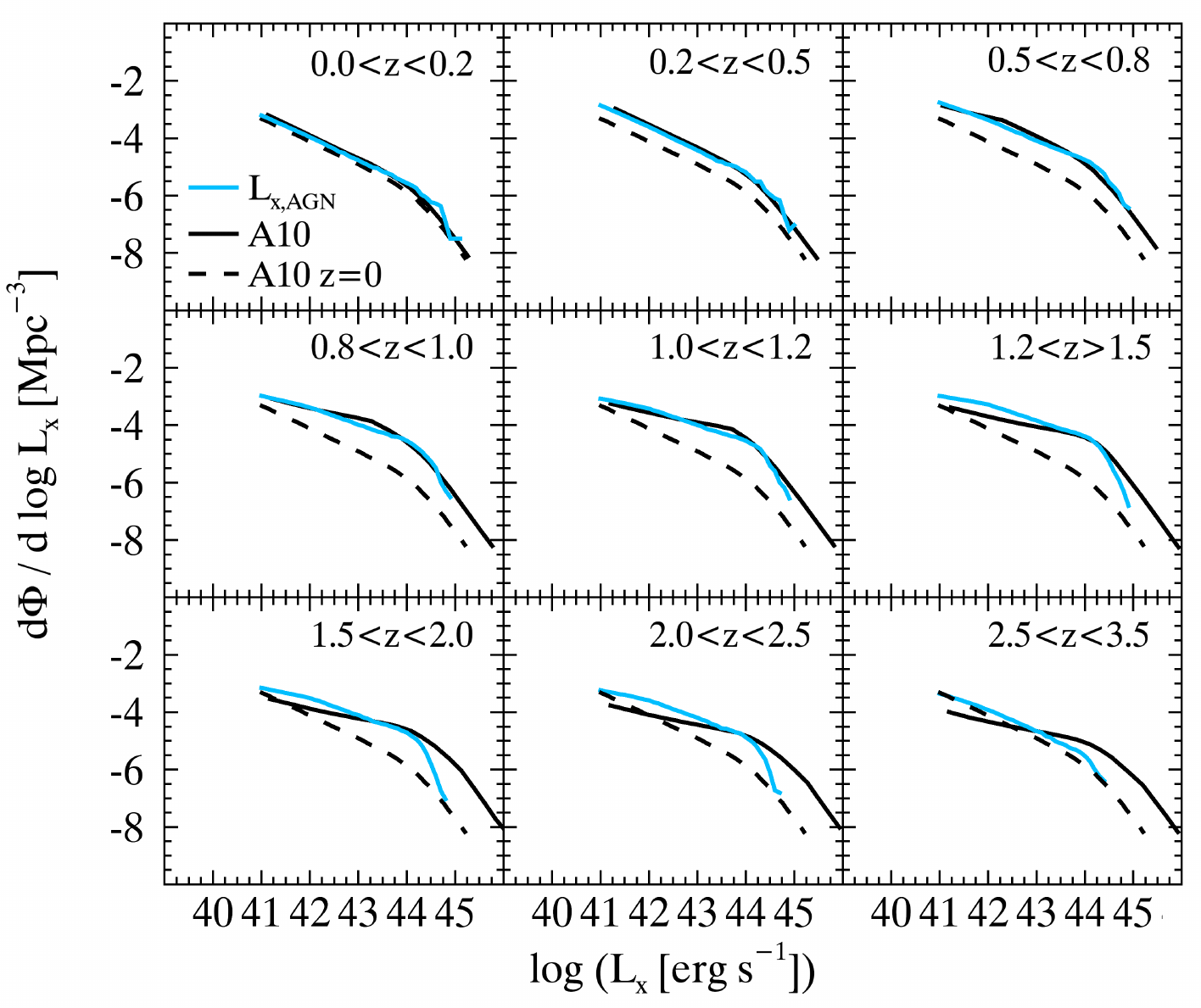}} \\
\caption{Comparison of the X-ray luminosity function (XLF) evolution from our model to results from X-ray observations. The XLF evaluated near the center of each redshift bin for the full sample of model galaxies is shown as a solid blue line. These model XLF are compared to the \citet{Air10} XLF for these same bins (solid black line) and the best Eddington distribution parameters are chosen by a minimum chi-squared calculation. The \citet{Air10} XLF for $z=0$ is shown in each bin for comparison. We find good agreement to $z\sim1.2$. Deviations at higher redshift are likely due to the simplicity of the model and may be solved by adding additional complexities, such as connecting AGN accretion to the star formation rate. \label{fig:evolxlf}}
\end{center}
\end{figure*}

We compare our model initially to the luminosity-dependent density evolution (LDDE) model of the hard X-ray luminosity function from \citet{Air10}. We are able to ``tune'' our model to match the \citet{Air10} XLF at $z=0$ by varying our model AGN Eddington ratio distribution $\alpha$ and lower cutoff parameters. We approximately match the distribution with $\alpha=0.8$ and $\text{L}_{cut}=-6.0$ $\text{L}_{bol}/\text{L}_{Edd}$. This Schechter function distribution deviates from the parameters presented in \citet{Jon16} ($\alpha=0.4$ and $\text{L}_{cut}=-3.75$ $\text{L}_{bol}/\text{L}_{Edd}$). We expect that alpha will be steeper since our distribution is describing AGN activity in both the active and quiescent galaxy population \citep{Gab13}, whereas in \citet{Jon16}, we focus on the Eddington ratio distribution of star forming galaxies alone. It is worthwhile to note that when comparing these theoretical distributions at $z=0$ with observations that are higher redshifts, an exact comparison is difficult due to the evolution of the XLF (e.g., \citealt{Ued14}).

The luminosity function that we calculate with our simulated AGN population appears as a double power law with its break at $\text{L}_X\sim10^{44}\;\text{erg s}^{-1}$ (solid grey line, Figure \ref{fig:lfn}). This characteristic shape is influenced by the black hole mass function and is consistent with what is modeled by \citet[][orange, Figure \ref{fig:lfn}]{Air10}. At low luminosities we observe a turnover due to the galaxy dominating the X-ray emission and the minimum contribution from star formation as imposed by our black hole mass limit. Since our simulation relies on the random selection of the accretion rate from the Eddington ratio distribution, as well as randomly assigning which galaxies are obscured, we ran a bootstrap resampling to demonstrate the variance of our simulation due to these random selections. We show these random assignments and limited volume do not introduce a significant uncertainty to our model XLF (grey area, Figure \ref{fig:lfn}).

We can further examine the effects of selecting AGN based on two different thresholds on the ``observed'' XLF. The Eddington ratio limited sample (solid purple line, Figure \ref{fig:lfn}) yields a similar wide distribution in luminosity ($39.5\lesssim\log\text{L}_X\lesssim45.5$) compared to the full model galaxies (solid grey line, Figure \ref{fig:lfn}). We note that in practice, it is difficult to select AGN below $\text{L}_X \sim10^{41}\;\text{erg s}^{-1}$ due to contamination from other accretion processes (e.g., \citealt{Leh16}). Imposing a luminosity limit (solid green line, Figure \ref{fig:lfn}), as we would expect, causes the XLF to truncate below $\text{L}_X=10^{41.5}\;\text{erg s}^{-1}$ as it is defined, which is consistent with the LDDE model of \citet{Air10}. At high luminosities ($\text{L}_X\gtrsim10^{44}\;\text{erg s}^{-1}$), where all AGN are accreting at high Eddington rates, we find that our models, including those that have imposed limits, are consistent with the \citet{Air10} XLF at $z=0$. This suggests that our simple prescription for AGN accretion is valid to first order for describing the full AGN population at $z=0$.

\subsubsection{An evolving AGN luminosity function}

We further explore the evolution with redshift of the XLF in our model. AGN X-ray luminosities for galaxies in earlier (higher-z) snapshots from the \citet{Mut13} simulation are calculated following the $z=0$ prescription outlined in Section \ref{ssec:agn} (also using an Eddington ratio distribution given by a Schechter function). We calculate the minimum chi-squared fit between our model XLF and the \citet{Air10} AGN XLF for a selection of nine redshift bins in order to determine our best fit parameters for the Schechter function $\alpha$, and lower cutoff (Figure \ref{fig:evolxlf}). Over this redshift range, $\alpha$ varies between $0.3$ and $0.8$, while the lower cutoff varies between $-5.0$ and $-7.0$. We find that our XLF are qualitatively consistent with the \citet{Air10} XLF within these redshift bins to $z\sim1.2$ for our best fit parameters. This confirms the applicability of the model at $z\lesssim1$ for further comparisons to observed sSFR, mass, and halo occupation distribution results.

\subsection{Host Galaxy SFR and Mass}\label{ssec:color}

\begin{figure}[!t]
\resizebox{85mm}{!}{\includegraphics{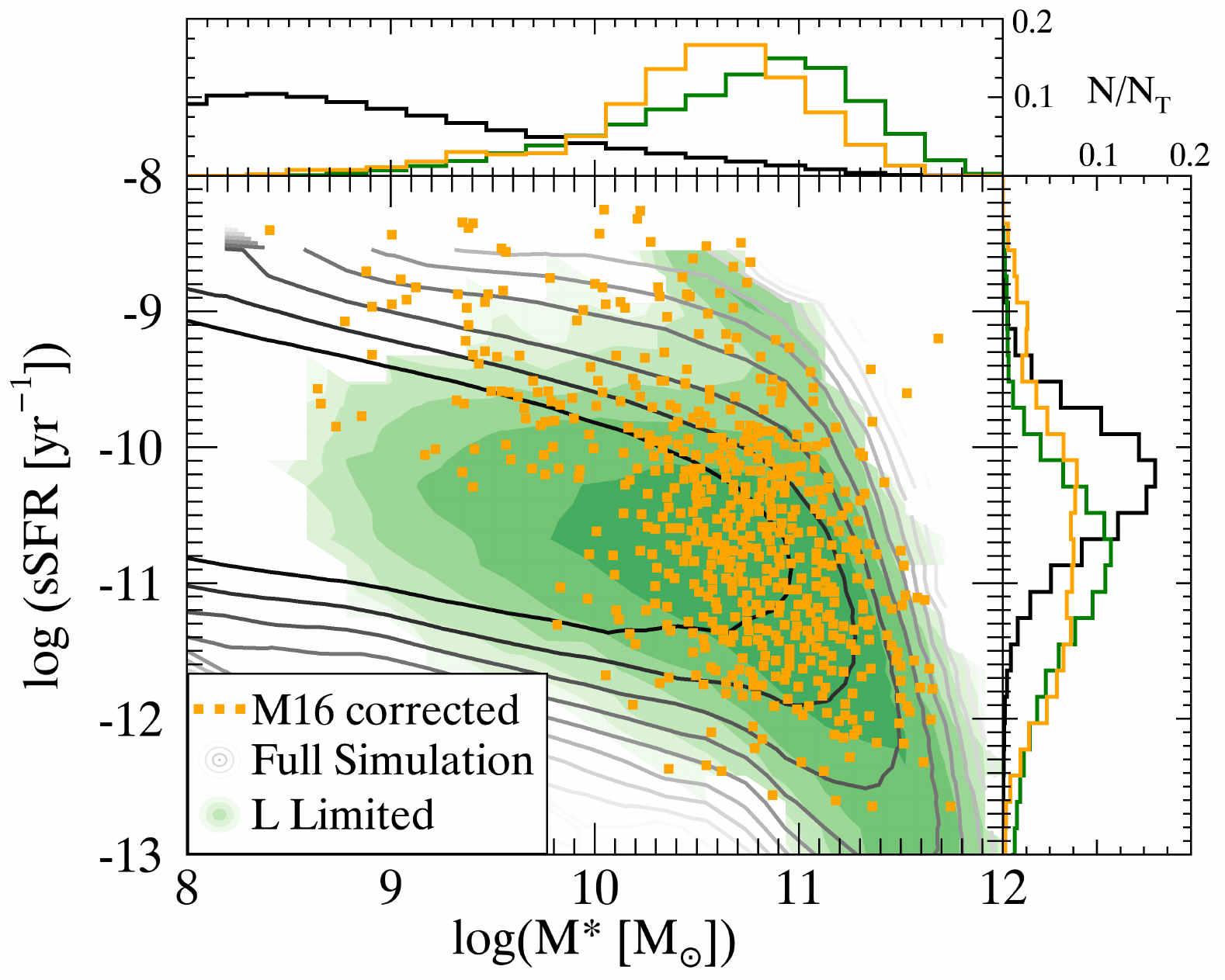}} \\
\caption{Specific star formation rate versus stellar mass distribution for both the full sample of model galaxies (grey-black) compared to the distribution for the model luminosity limited sample with lower limit of $\text{L}_{X}=10^{41.5}\;\text{erg s}^{-1}$. We do not include the sSFR-M* distribution of the Eddington limited sample for clarity since our Eddington ratios are tied to the stellar mass such that this limit does not change the overall shape of the full distribution. The corrected \citet[M16,][]{Men16} observed X-ray AGN are shown in orange squares and are consistently distributed with respect to the luminosity limited sample. Normalized histograms of each axis are shown for clarity.\label{fig:color}}
\end{figure}

The \citet{Mut13} simulation tracks the star formation rate of our model galaxies, thus we can calculate the specific star formation rate (sSFR; ratio of the star formation rate to the stellar mass). The full simulated distribution in sSFR-stellar mass (grey contours, Figure \ref{fig:color}) is consistent with the shape of the \textsc{eagle} hydrodynamical theoretical distribution for $z=0$ \citep{Guo16} as well as the observed sSFR-stellar mass distributions from \citet{Sch07} ($0.01<z<0.25$) and the GAMA survey \citep[][$0.05<z<0.32$]{Bau13}. Thus, with our simple prescription, we are able to match more complex models to first order.

By design, our Eddington ratios are tied directly to the stellar mass through our black hole mass calculations. Thus, applying our Eddington ratio cut to the full sample does not change the overall shape of the distribution in sSFR-stellar mass but simply limits the total number of AGN selected in the sample.

We now apply a luminosity limit to ``define'' our AGN and compare the resulting distributions to the X-ray observations from \citet{Men16}. \citet{Men16} utilizes an X-ray luminosity threshold of $\text{L}_X=10^{41.5}\;\text{erg s}^{-1}$ to select AGN from $0.2<z<1.2$. In order to more accurately compare these results to our model output at $z=0$, we correct the sSFR of this observed comparison sample using the \citet{Lee15} galaxy main sequence evolution ($\text{SFR}_0\propto(1+z)^{4.12\pm0.08}$) at $z=0.2$. We find comparable ``observed'' distributions that are truncated at higher stellar mass compared to our intrinsic AGN population; we are able to match observations that are different from our intrinsic distribution by using appropriate limits. We see a slight shift of our luminosity limited model to higher stellar masses compared to the Mendez sample as shown by the histograms. 

Our overestimation of the stellar masses may in part be due to the connection between AGN accretion and star formation, such that the AGN fraction decreases for passive galaxies (e.g., \citealt{Che13,Aza15}). AGN suppression in high mass systems is not yet included in this simple model.

With the addition of the galaxy main sequence evolution to the \citet{Men16} sample, our limited model sSFR appears consistent with observations. Using a straightforward and simple semi-numerical model, we are able to build comparable intrinsic sSFR-stellar mass distributions, as well as broadly match what is observed with appropriate limits.

\subsection{The Halo Occupation Distribution of AGN}\label{ssec:hod}

\begin{figure}[!t]
\resizebox{85mm}{!}{\includegraphics{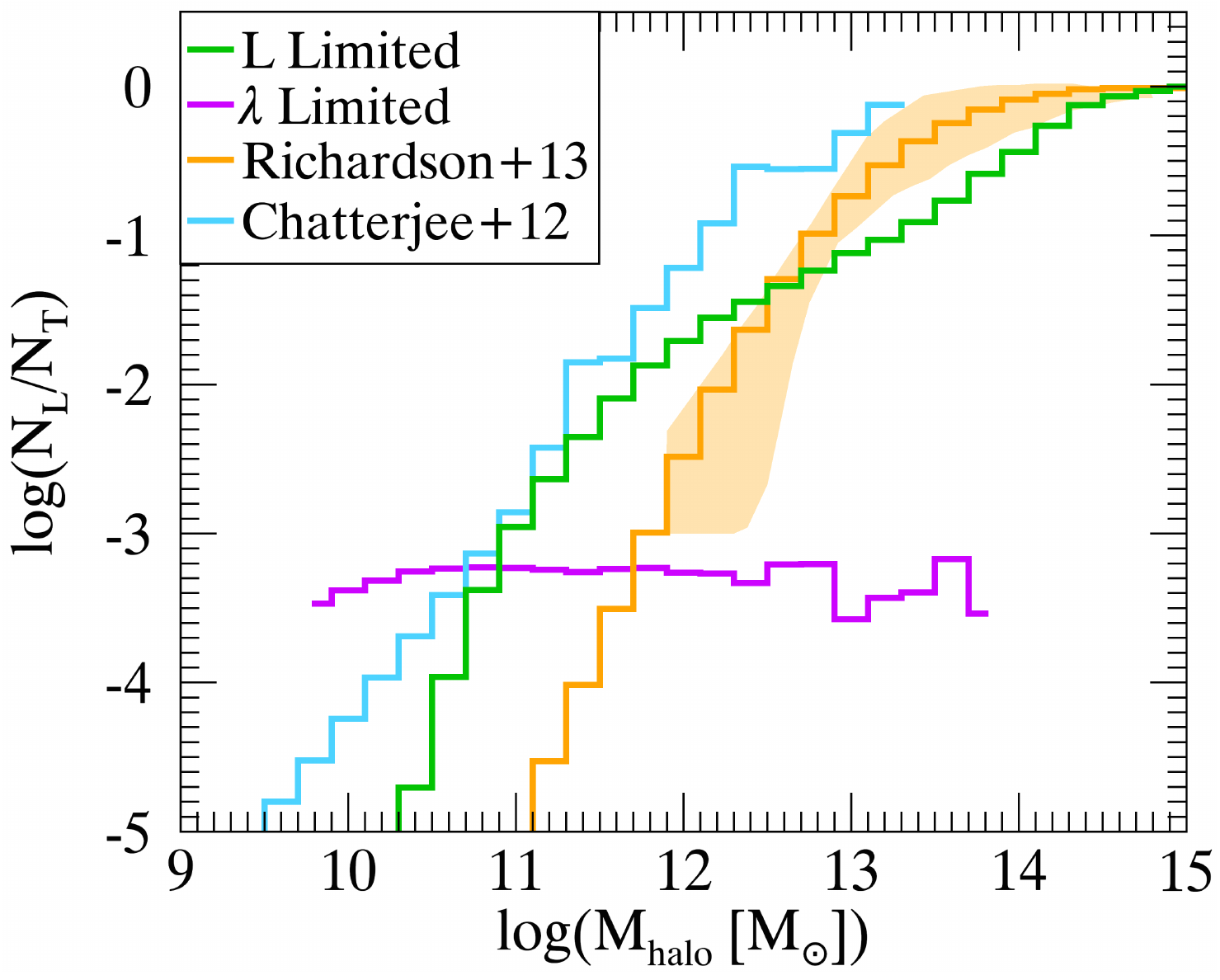}} \\
\caption{The mean halo occupation distribution (HOD) of both the luminosity limited, $\text{L}_{X}>10^{41.5}\;\text{erg s}^{-1}$, (green) and Eddington limited, $\lambda>0.01$, (magenta) samples of simulated AGN. We compare our modeled samples to the observational HOD analysis of \citet{Ric13} at $z\sim1.2$ (orange) and the theoretically motivated HOD of \citet{Cha12} at $z=1.0$ with luminosity limit $\text{L}_{X}>10^{42}\;\text{erg s}^{-1}$ (cyan). While not able to be compared directly, the luminosity limited model HOD at $z=0$ lies within the overall shape of the \citet{Ric13} and \citet{Cha12} HOD.\label{fig:hod}}
\end{figure}

We further investigate the host dark matter halos of our simulated AGN and compare our simple model to results from the observational clustering analysis of \citet{Ric13} as well as the simulation of \citet{Cha12}. We compute the halo occupation distribution (HOD) for luminosity and Eddington ratio limited samples with respect to our intrinsic full sample. As shown in the color-mass analysis, our Eddington ratio limited sample (magenta, Figure \ref{fig:hod}) follows a distribution consistent with the intrinsic sample and thus the distribution of the ratio appears flat on the HOD. This suggests that by selecting AGN based on Eddington ratio we are capturing the black hole growth across all mass limits.

Our luminosity limited sample, however, deviates from our model intrinsic sample so the HOD takes on a positive slope with a turnover at $M\sim10^{12} M_\odot$. We compare this shape to the \citet{Ric13} distribution at $z\sim1.2$ and find a similar turnover, albeit at higher halo masses. The \citet{Cha12} HOD at $z=1.0$ similarly exhibits a positive slope. While these two HODs represent populations at varying halo mass and differing redshifts, they provide a general shape and range with which we may check the validity of our local universe luminosity limited sample distribution. We find that our model HOD with luminosity limits built from a simple semi-numerical model attains the general shape of a theoretically motivated HOD as well as the observationally informed HOD. This reinforces the result that a simple prescription can describe the observed AGN population to first order and by varying the criteria that define our AGN population, we are selecting AGN with different host galaxy and halo properties.

\section{Summary}\label{sec:sum}

Our goal has been to build a simulation of the full AGN population and their host galaxy properties using a straightforward semi-numerical model, in order to investigate the impact of AGN selection criteria on the ``observed'' population. We begin with a sample of dark matter halos from the Millennium N-body simulation \citep{Spr05} and connect galactic growth to the dark matter halo formation history using the \citet{Mut13} formation history semi-numerical model. Our AGN accretion model is motivated by the results of \citet{Jon16} and is straightforward and computationally inexpensive. It treats AGN accretion as an instantaneous rate selected from an Eddington ratio distribution that takes the functional form of a Schechter function. Using a variety of galaxy and AGN scaling relationships, we convert our simulated bolometric luminosities into X-ray luminosities to better compare to observations. We assign AGN to be obscured or unobscured following the prescription outlined in Section \ref{ssec:agn}.

With our simulated full AGN population, we explicitly model selection effects by defining our ``observed'' AGN samples based on typical observational limits; a luminosity threshold of $\text{L}_{X}>10^{41.5}\;\text{erg s}^{-1}$ and an Eddington ratio limit of $\lambda\gtrsim0.01$. We then compare our intrinsic AGN population, as well as the two limited samples, to a variety of published relationships. 
We are able to recreate the \citet{Air10} XLF using a Schechter function Eddington ratio distribution with a power-law slope of $\alpha=0.8$ and lower cutoff of $\text{L}_{cut}=-6.0$. 
The sSFR-M* distribution of our intrinsic AGN population is broadly consistent with that of the \textsc{eagle} hydrodynamical simulation at $z=0$ \citep{Guo16} as well as the observational distributions of \citet{Sch07} ($0.01<x<0.25$) and \citet{Bau13} ($0.05<z<0.32$), while our luminosity limited AGN population matches the observed X-ray AGN of \citet{Men16}. 
We further examine the consequences of these selection criteria by comparing our model HODs to the observational analysis from \citet{Ric13} and theoretical model of \citet{Cha12} and find that our luminosity limited sample yields the same general shape of the HOD, while our Eddington limited sample maintains the distribution of our intrinsic sample. 
We find that our straightforward prescription for AGN accretion is able to recreate both theoretical and observational relationships for the ``observed'' AGN population, once AGN selection criteria are taken into account. 

AGN are often defined using a luminosity or Eddington threshold. We have shown that each criterion selects different ``observed'' AGN properties and host galaxy parameters from our intrinsic full AGN population. Selecting AGN based on an Eddington ratio limit yields a wider range of AGN and host galaxy properties compared to a luminosity limit and may better represent the underlying AGN population. A luminosity limited sample is often driven by the sensitivity of surveys; our luminosity limited model deviates most from the intrinsic full population. This selection effect can be partially mitigated by selecting AGN properties that are directly related to the AGN accretion rate. This work highlights the difficulty in correcting for even the most simple of selection effects in AGN X-ray observations.

In the future, this analysis can be extended to higher redshifts and tied to star formation using the full history of the \citet{Mut13} model and our simple prescription for AGN accretion. Modeling the AGN with additional selection techniques such as luminosity thresholds at various wavelengths and photometric color selection will be particularly useful for comparison with survey data.

\acknowledgments
We thank our collaborators as well as the referee for constructive comments that improved the paper. This work was supported in part by the National Aeronautics and Space Administration under Grant Number NNX15AU32H issued through the NASA Education Minority University Research Education Project (MUREP) through the NASA Harriett G. Jenkins Graduate Fellowship activity. This research has made use of NASA's Astrophysics Data System.


\bibliography{ap_select}

\begin{thebibliography}{}
\expandafter\ifx\csname natexlab\endcsname\relax\def\natexlab#1{#1}\fi
\providecommand{\url}[1]{\href{#1}{#1}}

\bibitem[{{Aird} {et~al.}(2010)}]{Air10}
{Aird}, J., {et~al.} 2010, MNRAS, 401, 2531

\bibitem[{{Aird} {et~al.}(2012){Aird}, {Coil}, {Moustakas}, {Blanton},
  {Burles}, {Cool}, {Eisenstein}, {Smith}, {Wong}, \& {Zhu}}]{Air12}
{Aird}, J., {Coil}, A.~L., {Moustakas}, J., {et~al.} 2012, \apj, 746, 90

\bibitem[{{Alexander} \& {Hickox}(2012)}]{Ale12}
{Alexander}, D.~M., \& {Hickox}, R.~C. 2012, NAR, 56, 93

\bibitem[{{Azadi} {et~al.}(2015){Azadi}, {Aird}, {Coil}, {Moustakas}, {Mendez},
  {Blanton}, {Cool}, {Eisenstein}, {Wong}, \& {Zhu}}]{Aza15}
{Azadi}, M., {Aird}, J., {Coil}, A.~L., {et~al.} 2015, \apj, 806, 187

\bibitem[{{Bauer} {et~al.}(2013){Bauer}, {Hopkins}, {Gunawardhana}, {Taylor},
  {Baldry}, {Bamford}, {Bland-Hawthorn}, {Brough}, {Brown}, {Cluver},
  {Colless}, {Conselice}, {Croom}, {Driver}, {Foster}, {Jones}, {Lara-Lopez},
  {Liske}, {L{\'o}pez-S{\'a}nchez}, {Loveday}, {Norberg}, {Owers}, {Pimbblet},
  {Robotham}, {Sansom}, \& {Sharp}}]{Bau13}
{Bauer}, A.~E., {Hopkins}, A.~M., {Gunawardhana}, M., {et~al.} 2013, \mnras,
  434, 209

\bibitem[{{Bennert} {et~al.}(2011){Bennert}, {Auger}, {Treu}, {Woo}, \&
  {Malkan}}]{Ben11}
{Bennert}, V.~N., {Auger}, M.~W., {Treu}, T., {Woo}, J.-H., \& {Malkan}, M.~A.
  2011, \apj, 742, 107

\bibitem[{{B{\'e}thermin} {et~al.}(2012){B{\'e}thermin}, {Dor{\'e}}, \&
  {Lagache}}]{Bet12}
{B{\'e}thermin}, M., {Dor{\'e}}, O., \& {Lagache}, G. 2012, \aap, 537, L5

\bibitem[{{Bongiorno} {et~al.}(2012){Bongiorno}, {Merloni}, {Brusa},
  {Magnelli}, {Salvato}, {Mignoli}, {Zamorani}, {Fiore}, {Rosario}, {Mainieri},
  {Hao}, {Comastri}, {Vignali}, {Balestra}, {Bardelli}, {Berta}, {Civano},
  {Kampczyk}, {Le Floc'h}, {Lusso}, {Lutz}, {Pozzetti}, {Pozzi}, {Riguccini},
  {Shankar}, \& {Silverman}}]{Bon12}
{Bongiorno}, A., {Merloni}, A., {Brusa}, M., {et~al.} 2012, \mnras, 427, 3103

\bibitem[{{Bower} {et~al.}(2006){Bower}, {Benson}, {Malbon}, {Helly}, {Frenk},
  {Baugh}, {Cole}, \& {Lacey}}]{Bow06}
{Bower}, R.~G., {Benson}, A.~J., {Malbon}, R., {et~al.} 2006, \mnras, 370, 645

\bibitem[{{Chatterjee} {et~al.}(2012){Chatterjee}, {Degraf}, {Richardson},
  {Zheng}, {Nagai}, \& {Di Matteo}}]{Cha12}
{Chatterjee}, S., {Degraf}, C., {Richardson}, J., {et~al.} 2012, \mnras, 419,
  2657

\bibitem[{{Chen} {et~al.}(2013){Chen}, {Hickox}, {Alberts}, {Brodwin}, {Jones},
  {Murray}, {Alexander}, {Assef}, {Brown}, {Dey}, {Forman}, {Gorjian},
  {Goulding}, {Le Floc'h}, {Jannuzi}, {Mullaney}, \& {Pope}}]{Che13}
{Chen}, C.-T.~J., {Hickox}, R.~C., {Alberts}, S., {et~al.} 2013, \apj, 773, 3

\bibitem[{{Conroy} \& {Wechsler}(2009)}]{Con09}
{Conroy}, C., \& {Wechsler}, R.~H. 2009, \apj, 696, 620

\bibitem[{{Croton} {et~al.}(2006){Croton}, {Springel}, {White}, {De Lucia},
  {Frenk}, {Gao}, {Jenkins}, {Kauffmann}, {Navarro}, \& {Yoshida}}]{Cro06}
{Croton}, D.~J., {Springel}, V., {White}, S.~D.~M., {et~al.} 2006, \mnras, 365,
  11

\bibitem[{{Decarli} {et~al.}(2010){Decarli}, {Falomo}, {Treves}, {Labita},
  {Kotilainen}, \& {Scarpa}}]{Dec10}
{Decarli}, R., {Falomo}, R., {Treves}, A., {et~al.} 2010, \mnras, 402, 2453

\bibitem[{{DeGraf} {et~al.}(2015){DeGraf}, {Di Matteo}, {Treu}, {Feng}, {Woo},
  \& {Park}}]{Deg15}
{DeGraf}, C., {Di Matteo}, T., {Treu}, T., {et~al.} 2015, \mnras, 454, 913

\bibitem[{{Feng} {et~al.}(2016){Feng}, {Di-Matteo}, {Croft}, {Bird},
  {Battaglia}, \& {Wilkins}}]{Fen16}
{Feng}, Y., {Di-Matteo}, T., {Croft}, R.~A., {et~al.} 2016, \mnras, 455, 2778

\bibitem[{{Fragos} {et~al.}(2013)}]{Fra13}
{Fragos}, T., {et~al.} 2013, ApJ, 764, 41

\bibitem[{{Gabor} \& {Bournaud}(2013)}]{Gab13}
{Gabor}, J.~M., \& {Bournaud}, F. 2013, \mnras, 434, 606

\bibitem[{{Genel} {et~al.}(2014){Genel}, {Vogelsberger}, {Springel}, {Sijacki},
  {Nelson}, {Snyder}, {Rodriguez-Gomez}, {Torrey}, \& {Hernquist}}]{Gen14}
{Genel}, S., {Vogelsberger}, M., {Springel}, V., {et~al.} 2014, \mnras, 445,
  175

\bibitem[{{Guo} {et~al.}(2016){Guo}, {Gonzalez-Perez}, {Guo}, {Schaller},
  {Furlong}, {Bower}, {Cole}, {Crain}, {Frenk}, {Helly}, {Lacey}, {Lagos},
  {Mitchell}, {Schaye}, \& {Theuns}}]{Guo16}
{Guo}, Q., {Gonzalez-Perez}, V., {Guo}, Q., {et~al.} 2016, \mnras, 461, 3457

\bibitem[{{H{\"a}ring} \& {Rix}(2004)}]{Har04}
{H{\"a}ring}, N., \& {Rix}, H.-W. 2004, \apjl, 604, L89

\bibitem[{{Heckman} {et~al.}(2004){Heckman}, {Kauffmann}, {Brinchmann},
  {Charlot}, {Tremonti}, \& {White}}]{Hec04}
{Heckman}, T.~M., {Kauffmann}, G., {Brinchmann}, J., {et~al.} 2004, \apj, 613,
  109

\bibitem[{{Henriques} {et~al.}(2009){Henriques}, {Thomas}, {Oliver}, \&
  {Roseboom}}]{Hen09}
{Henriques}, B.~M.~B., {Thomas}, P.~A., {Oliver}, S., \& {Roseboom}, I. 2009,
  \mnras, 396, 535

\bibitem[{{Hickox} {et~al.}(2014){Hickox}, {Mullaney}, {Alexander}, {Chen},
  {Civano}, {Goulding}, \& {Hainline}}]{Hic14}
{Hickox}, R.~C., {Mullaney}, J.~R., {Alexander}, D.~M., {et~al.} 2014, \apj,
  782, 9

\bibitem[{{Hickox} {et~al.}(2009){Hickox}, {Jones}, {Forman}, {Murray},
  {Kochanek}, {Eisenstein}, {Jannuzi}, {Dey}, {Brown}, {Stern}, {Eisenhardt},
  {Gorjian}, {Brodwin}, {Narayan}, {Cool}, {Kenter}, {Caldwell}, \&
  {Anderson}}]{Hic09}
{Hickox}, R.~C., {Jones}, C., {Forman}, W.~R., {et~al.} 2009, \apj, 696, 891

\bibitem[{{Hopkins} {et~al.}(2005){Hopkins}, {Hernquist}, {Cox}, {Di Matteo},
  {Robertson}, \& {Springel}}]{Hop05}
{Hopkins}, P.~F., {Hernquist}, L., {Cox}, T.~J., {et~al.} 2005, \apj, 630, 716

\bibitem[{{Hopkins} {et~al.}(2009){Hopkins}, {Hickox}, {Quataert}, \&
  {Hernquist}}]{Hop09}
{Hopkins}, P.~F., {Hickox}, R., {Quataert}, E., \& {Hernquist}, L. 2009,
  \mnras, 398, 333

\bibitem[{{Hornschemeier} {et~al.}(2005)}]{Hor05}
{Hornschemeier}, A.~E., {et~al.} 2005, AJ, 129, 86

\bibitem[{{Janiuk} \& {Czerny}(2011)}]{Jan11}
{Janiuk}, A., \& {Czerny}, B. 2011, \mnras, 414, 2186

\bibitem[{{Jones} {et~al.}(2016){Jones}, {Hickox}, {Black}, {Hainline},
  {DiPompeo}, \& {Goulding}}]{Jon16}
{Jones}, M.~L., {Hickox}, R.~C., {Black}, C.~S., {et~al.} 2016, \apj, 826, 12

\bibitem[{Kauffmann \& Heckman(2009)}]{KH09}
Kauffmann, G., \& Heckman, T.~M. 2009, MNRAS, 397, 135

\bibitem[{{Kauffmann} {et~al.}(1993){Kauffmann}, {White}, \&
  {Guiderdoni}}]{Kau93}
{Kauffmann}, G., {White}, S.~D.~M., \& {Guiderdoni}, B. 1993, \mnras, 264, 201

\bibitem[{{Keel} {et~al.}(2012){Keel}, {Chojnowski}, {Bennert}, {Schawinski},
  {Lintott}, {Lynn}, {Pancoast}, {Harris}, {Nierenberg}, {Sonnenfeld}, \&
  {Proctor}}]{Kee12}
{Keel}, W.~C., {Chojnowski}, S.~D., {Bennert}, V.~N., {et~al.} 2012, \mnras,
  420, 878

\bibitem[{{Khandai} {et~al.}(2015){Khandai}, {Di Matteo}, {Croft}, {Wilkins},
  {Feng}, {Tucker}, {DeGraf}, \& {Liu}}]{Kha15}
{Khandai}, N., {Di Matteo}, T., {Croft}, R., {et~al.} 2015, \mnras, 450, 1349

\bibitem[{{Kormendy} \& {Ho}(2013)}]{Kor13}
{Kormendy}, J., \& {Ho}, L.~C. 2013, ARAA, 51, 511

\bibitem[{{Koutoulidis} {et~al.}(2013){Koutoulidis}, {Plionis},
  {Georgantopoulos}, \& {Fanidakis}}]{Kou13}
{Koutoulidis}, L., {Plionis}, M., {Georgantopoulos}, I., \& {Fanidakis}, N.
  2013, \mnras, 428, 1382

\bibitem[{{Lansbury} {et~al.}(2014){Lansbury}, {Alexander}, {Del Moro},
  {Gandhi}, {Assef}, {Stern}, {Aird}, {Ballantyne}, {Balokovi{\'c}}, {Bauer},
  {Boggs}, {Brandt}, {Christensen}, {Craig}, {Elvis}, {Grefenstette}, {Hailey},
  {Harrison}, {Hickox}, {Koss}, {LaMassa}, {Luo}, {Mullaney}, {Teng}, {Urry},
  \& {Zhang}}]{Lan14}
{Lansbury}, G.~B., {Alexander}, D.~M., {Del Moro}, A., {et~al.} 2014, \apj,
  785, 17

\bibitem[{{Lansbury} {et~al.}(2015){Lansbury}, {Gandhi}, {Alexander}, {Assef},
  {Aird}, {Annuar}, {Ballantyne}, {Balokovi{\'c}}, {Bauer}, {Boggs}, {Brandt},
  {Brightman}, {Christensen}, {Civano}, {Comastri}, {Craig}, {Del Moro},
  {Grefenstette}, {Hailey}, {Harrison}, {Hickox}, {Koss}, {LaMassa}, {Luo},
  {Puccetti}, {Stern}, {Treister}, {Vignali}, {Zappacosta}, \& {Zhang}}]{Lan15}
{Lansbury}, G.~B., {Gandhi}, P., {Alexander}, D.~M., {et~al.} 2015, \apj, 809,
  115

\bibitem[{{Lee} {et~al.}(2015){Lee}, {Sanders}, {Casey}, {Toft}, {Scoville},
  {Hung}, {Le Floc'h}, {Ilbert}, {Zahid}, {Aussel}, {Capak}, {Kartaltepe},
  {Kewley}, {Li}, {Schawinski}, {Sheth}, \& {Xiao}}]{Lee15}
{Lee}, N., {Sanders}, D.~B., {Casey}, C.~M., {et~al.} 2015, \apj, 801, 80

\bibitem[{{Lehmer} {et~al.}(2010){Lehmer}, {Alexander}, {Bauer}, {Brandt},
  {Goulding}, {Jenkins}, {Ptak}, \& {Roberts}}]{Leh10}
{Lehmer}, B.~D., {Alexander}, D.~M., {Bauer}, F.~E., {et~al.} 2010, \apj, 724,
  559

\bibitem[{{Lehmer} {et~al.}(2016){Lehmer}, {Basu-Zych}, {Mineo}, {Brandt},
  {Eufrasio}, {Fragos}, {Hornschemeier}, {Luo}, {Xue}, {Bauer}, {Gilfanov},
  {Ranalli}, {Schneider}, {Shemmer}, {Tozzi}, {Trump}, {Vignali}, {Wang},
  {Yukita}, \& {Zezas}}]{Leh16}
{Lehmer}, B.~D., {Basu-Zych}, A.~R., {Mineo}, S., {et~al.} 2016, \apj, 825, 7

\bibitem[{{Lu} {et~al.}(2011){Lu}, {Mo}, {Weinberg}, \& {Katz}}]{Lu11}
{Lu}, Y., {Mo}, H.~J., {Weinberg}, M.~D., \& {Katz}, N. 2011, \mnras, 416, 1949

\bibitem[{{Lusso} {et~al.}(2012){Lusso}, {Comastri}, {Simmons}, {Mignoli},
  {Zamorani}, {Vignali}, {Brusa}, {Shankar}, {Lutz}, {Trump}, {Maiolino},
  {Gilli}, {Bolzonella}, {Puccetti}, {Salvato}, {Impey}, {Civano}, {Elvis},
  {Mainieri}, {Silverman}, {Koekemoer}, {Bongiorno}, {Merloni}, {Berta}, {Le
  Floc'h}, {Magnelli}, {Pozzi}, \& {Riguccini}}]{Lus12}
{Lusso}, E., {Comastri}, A., {Simmons}, B.~D., {et~al.} 2012, \mnras, 425, 623

\bibitem[{{Mendez} {et~al.}(2016){Mendez}, {Coil}, {Aird}, {Skibba},
  {Diamond-Stanic}, {Moustakas}, {Blanton}, {Cool}, {Eisenstein}, {Wong}, \&
  {Zhu}}]{Men16}
{Mendez}, A.~J., {Coil}, A.~L., {Aird}, J., {et~al.} 2016, \apj, 821, 55

\bibitem[{{Merloni} {et~al.}(2010){Merloni}, {Bongiorno}, {Bolzonella},
  {Brusa}, {Civano}, {Comastri}, {Elvis}, {Fiore}, {Gilli}, {Hao}, {Jahnke},
  {Koekemoer}, {Lusso}, {Mainieri}, {Mignoli}, {Miyaji}, {Renzini}, {Salvato},
  {Silverman}, {Trump}, {Vignali}, {Zamorani}, {Capak}, {Lilly}, {Sanders},
  {Taniguchi}, {Bardelli}, {Carollo}, {Caputi}, {Contini}, {Coppa}, {Cucciati},
  {de la Torre}, {de Ravel}, {Franzetti}, {Garilli}, {Hasinger}, {Impey},
  {Iovino}, {Iwasawa}, {Kampczyk}, {Kneib}, {Knobel}, {Kova{\v c}},
  {Lamareille}, {Le Borgne}, {Le Brun}, {Le F{\`e}vre}, {Maier}, {Pello},
  {Peng}, {Perez Montero}, {Ricciardelli}, {Scodeggio}, {Tanaka}, {Tasca},
  {Tresse}, {Vergani}, \& {Zucca}}]{Mer10}
{Merloni}, A., {Bongiorno}, A., {Bolzonella}, M., {et~al.} 2010, \apj, 708, 137

\bibitem[{{Merloni} {et~al.}(2014){Merloni}, {Bongiorno}, {Brusa}, {Iwasawa},
  {Mainieri}, {Magnelli}, {Salvato}, {Berta}, {Cappelluti}, {Comastri},
  {Fiore}, {Gilli}, {Koekemoer}, {Le Floc'h}, {Lusso}, {Lutz}, {Miyaji},
  {Pozzi}, {Riguccini}, {Rosario}, {Silverman}, {Symeonidis}, {Treister},
  {Vignali}, \& {Zamorani}}]{Mer14}
{Merloni}, A., {Bongiorno}, A., {Brusa}, M., {et~al.} 2014, \mnras, 437, 3550

\bibitem[{{Mutch} {et~al.}(2013{\natexlab{a}}){Mutch}, {Poole}, \&
  {Croton}}]{Mut13sam}
{Mutch}, S.~J., {Poole}, G.~B., \& {Croton}, D.~J. 2013{\natexlab{a}}, \mnras,
  428, 2001

\bibitem[{{Mutch} {et~al.}(2013{\natexlab{b}})}]{Mut13}
{Mutch}, S.~J., {et~al.} 2013{\natexlab{b}}, MNRAS, 435, 2445

\bibitem[{{Nandra} {et~al.}(2005){Nandra}, {Laird}, \& {Steidel}}]{Nan05}
{Nandra}, K., {Laird}, E.~S., \& {Steidel}, C.~C. 2005, \mnras, 360, L39

\bibitem[{{Novak} {et~al.}(2011){Novak}, {Ostriker}, \& {Ciotti}}]{Nov11}
{Novak}, G.~S., {Ostriker}, J.~P., \& {Ciotti}, L. 2011, \apj, 737, 26

\bibitem[{{Reddick} {et~al.}(2013){Reddick}, {Wechsler}, {Tinker}, \&
  {Behroozi}}]{Red13}
{Reddick}, R.~M., {Wechsler}, R.~H., {Tinker}, J.~L., \& {Behroozi}, P.~S.
  2013, \apj, 771, 30

\bibitem[{{Reines} \& {Volonteri}(2015)}]{Rei15}
{Reines}, A.~E., \& {Volonteri}, M. 2015, \apj, 813, 82

\bibitem[{{Richardson} {et~al.}(2013){Richardson}, {Chatterjee}, {Zheng},
  {Myers}, \& {Hickox}}]{Ric13}
{Richardson}, J., {Chatterjee}, S., {Zheng}, Z., {Myers}, A.~D., \& {Hickox},
  R. 2013, \apj, 774, 143

\bibitem[{{Sartori} {et~al.}(2016){Sartori}, {Schawinski}, {Koss}, {Treister},
  {Maksym}, {Keel}, {Urry}, {Lintott}, \& {Wong}}]{Sar16}
{Sartori}, L.~F., {Schawinski}, K., {Koss}, M., {et~al.} 2016, \mnras, 457,
  3629

\bibitem[{{Schawinski} {et~al.}(2015){Schawinski}, {Koss}, {Berney}, \&
  {Sartori}}]{Sch15}
{Schawinski}, K., {Koss}, M., {Berney}, S., \& {Sartori}, L.~F. 2015, \mnras,
  451, 2517

\bibitem[{{Schawinski} {et~al.}(2010){Schawinski}, {Evans}, {Virani}, {Urry},
  {Keel}, {Natarajan}, {Lintott}, {Manning}, {Coppi}, {Kaviraj}, {Bamford},
  {J{\'o}zsa}, {Garrett}, {van Arkel}, {Gay}, \& {Fortson}}]{Sch10qso}
{Schawinski}, K., {Evans}, D.~A., {Virani}, S., {et~al.} 2010, \apjl, 724, L30

\bibitem[{{Schaye} {et~al.}(2015){Schaye}, {Crain}, {Bower}, {Furlong},
  {Schaller}, {Theuns}, {Dalla Vecchia}, {Frenk}, {McCarthy}, {Helly},
  {Jenkins}, {Rosas-Guevara}, {White}, {Baes}, {Booth}, {Camps}, {Navarro},
  {Qu}, {Rahmati}, {Sawala}, {Thomas}, \& {Trayford}}]{EAGLE}
{Schaye}, J., {Crain}, R.~A., {Bower}, R.~G., {et~al.} 2015, \mnras, 446, 521

\bibitem[{{Schiminovich} {et~al.}(2007){Schiminovich}, {Wyder}, {Martin},
  {Johnson}, {Salim}, {Seibert}, {Treyer}, {Budav{\'a}ri}, {Hoopes},
  {Zamojski}, {Barlow}, {Forster}, {Friedman}, {Morrissey}, {Neff}, {Small},
  {Bianchi}, {Donas}, {Heckman}, {Lee}, {Madore}, {Milliard}, {Rich}, {Szalay},
  {Welsh}, \& {Yi}}]{Sch07}
{Schiminovich}, D., {Wyder}, T.~K., {Martin}, D.~C., {et~al.} 2007, \apjs, 173,
  315

\bibitem[{{Schirmer} {et~al.}(2013){Schirmer}, {Diaz}, {Holhjem}, {Levenson},
  \& {Winge}}]{Sch13}
{Schirmer}, M., {Diaz}, R., {Holhjem}, K., {Levenson}, N.~A., \& {Winge}, C.
  2013, \apj, 763, 60

\bibitem[{{Schulze} \& {Wisotzki}(2011)}]{Scz11}
{Schulze}, A., \& {Wisotzki}, L. 2011, \aap, 535, A87

\bibitem[{{Shankar} {et~al.}(2016){Shankar}, {Bernardi}, {Sheth}, {Ferrarese},
  {Graham}, {Savorgnan}, {Allevato}, {Marconi}, {L{\"a}sker}, \&
  {Lapi}}]{Sha16}
{Shankar}, F., {Bernardi}, M., {Sheth}, R.~K., {et~al.} 2016, \mnras, 460, 3119

\bibitem[{{Siemiginowska} \& {Elvis}(1997)}]{Sie97}
{Siemiginowska}, A., \& {Elvis}, M. 1997, \apjl, 482, L9

\bibitem[{{Silk} \& {Mamon}(2012)}]{Sil12}
{Silk}, J., \& {Mamon}, G.~A. 2012, Research in Astronomy and Astrophysics, 12,
  917

\bibitem[{{Spergel} {et~al.}(2003){Spergel}, {Verde}, {Peiris}, {Komatsu},
  {Nolta}, {Bennett}, {Halpern}, {Hinshaw}, {Jarosik}, {Kogut}, {Limon},
  {Meyer}, {Page}, {Tucker}, {Weiland}, {Wollack}, \& {Wright}}]{Spe03}
{Spergel}, D.~N., {Verde}, L., {Peiris}, H.~V., {et~al.} 2003, \apjs, 148, 175

\bibitem[{{Springel} {et~al.}(2005{\natexlab{a}}){Springel}, {Di Matteo}, \&
  {Hernquist}}]{Spr05feed}
{Springel}, V., {Di Matteo}, T., \& {Hernquist}, L. 2005{\natexlab{a}}, \mnras,
  361, 776

\bibitem[{{Springel} {et~al.}(2005{\natexlab{b}}){Springel}, {White},
  {Jenkins}, {Frenk}, {Yoshida}, {Gao}, {Navarro}, {Thacker}, {Croton},
  {Helly}, {Peacock}, {Cole}, {Thomas}, {Couchman}, {Evrard}, {Colberg}, \&
  {Pearce}}]{Spr05}
{Springel}, V., {White}, S.~D.~M., {Jenkins}, A., {et~al.} 2005{\natexlab{b}},
  \nat, 435, 629

\bibitem[{{Ueda} {et~al.}(2014){Ueda}, {Akiyama}, {Hasinger}, {Miyaji}, \&
  {Watson}}]{Ued14}
{Ueda}, Y., {Akiyama}, M., {Hasinger}, G., {Miyaji}, T., \& {Watson}, M.~G.
  2014, \apj, 786, 104

\bibitem[{{Veale} {et~al.}(2014){Veale}, {White}, \& {Conroy}}]{Vea14}
{Veale}, M., {White}, M., \& {Conroy}, C. 2014, \mnras, 445, 1144

\bibitem[{{Vogelsberger} {et~al.}(2014){Vogelsberger}, {Genel}, {Springel},
  {Torrey}, {Sijacki}, {Xu}, {Snyder}, {Nelson}, \& {Hernquist}}]{Vog14}
{Vogelsberger}, M., {Genel}, S., {Springel}, V., {et~al.} 2014, \mnras, 444,
  1518

\bibitem[{{Volonteri} {et~al.}(2015){Volonteri}, {Capelo}, {Netzer},
  {Bellovary}, {Dotti}, \& {Governato}}]{Vol15}
{Volonteri}, M., {Capelo}, P.~R., {Netzer}, H., {et~al.} 2015, \mnras, 449,
  1470

\bibitem[{{Woo} {et~al.}(2013){Woo}, {Schulze}, {Park}, {Kang}, {Kim}, \&
  {Riechers}}]{Woo13}
{Woo}, J.-H., {Schulze}, A., {Park}, D., {et~al.} 2013, \apj, 772, 49

\end{thebibliography}


\end{document}